# A Dielectric Metasurface Optical Chip for the Generation of Cold Atoms

Lingxiao Zhu[1,2#], Xuan Liu[3,4#], Basudeb Sain[5#], Mengyao Wang[1#], Christian Schlickriede[5], Yutao Tang[3], Junhong Deng[3,4], Kingfai Li[3], Jun Yang[2], Michael Holynski[1], Shuang Zhang[1], Thomas Zentgraf[5], Kai Bongs[1], Yu-Hung Lien[1]*, Guixin Li[3,4]*

[1]School of Physics and Astronomy, University of Birmingham, Birmingham, B15 2TT, UK.

[2]College of Intelligence Science and Technology, National University of Defense Technology, Changsha, 410073, China.

[3]Department of Materials Science and Engineering, Southern University of Science and Technology, Shenzhen 518055, China.

[4]Shenzhen Institute for Quantum Science and Engineering, Southern University of Science and Technology, Shenzhen 518055, China.

[5]Department of Physics, Paderborn University, Warburger Straße 100, 33098 Paderborn, Germany.

[#]These authors contributed equally to this work.
*Corresponding author. Email: y.lien@bham.ac.uk; ligx@sustech.edu.cn

**Abstract**
Compact and robust cold atom sources are increasingly important for quantum research, especially for transferring cutting-edge quantum science into practical applications. In this letter, we report on a novel scheme that utilizes a metasurface optical chip to replace the conventional bulky optical elements used to produce a cold atomic ensemble with a single incident laser beam, which is split by the metasurface into multiple beams of the desired polarization states. Atom numbers $\sim 10^7$ and temperatures (about 35 μK) of relevance to quantum sensing are achieved in a compact and robust fashion. Our work highlights the substantial progress towards fully integrated cold atom quantum devices by exploiting metasurface optical chips, which may have great potential in quantum sensing, quantum computing and other areas.

**Introduction**
Quantum systems based on cold atoms have enabled advances in areas such as quantum sensing (*1*), quantum metrology (*2, 3*) and quantum simulation (*4*). These cold atom quantum devices utilize light to engineer and interrogate the quantum states in accomplishing desired functionalities. The advances in photonic technologies greatly enhance the capabilities of the devices for controlling light and open up new horizons. One of the exciting developments is a metasurface optical device composed of spatially variant subwavelength structures, also called meta-atoms. They offer the capabilities of controlling the amplitude, polarization, and phase of light waves (*5–8*). Due to the versatility in the field of applications and design flexibility as well as straightforward fabrication methodology, metasurface-based optics can potentially replace or complement their conventional refractive and diffractive counterparts. The two-dimensional nature of metasurfaces opens the door of planar optics and many innovative planar optical elements have emerged ranging from linear to nonlinear optics, such as metalens (*9, 10*), optical holograms (*11, 12*), vortex beam generation (*13*), pulse shaping (*14*) and nonlinear optical phase and wavefront controlling (*15, 16*).

Recently, a number of studies demonstrated that metasurfaces have the enormous potential in quantum optics applications, such as the quantum metasurface interferometer, quantum entanglement states generation and reconstruction (*17–19*). To the best of our knowledge, metasurfaces have never been used to generate or manipulate cold atomic ensembles, which represents a highly resourceful quantum technology platform. Here, for the first time, we demonstrate a metasurface optical chip for the generation of cold atomic ensembles, providing a novel scheme for the realization of a single beam magneto-optical trap (MOT).

Cold atom quantum devices require preparation of cold atomic ensembles using laser cooling and trapping techniques for subsequent operation. Hot gas-phase atoms can be typically cooled and trapped using magneto-optical trapping (MOT), which combines laser cooling with a position dependent restoring force due to radiation pressure (*20, 21*). The standard MOT apparatus commonly utilizes three orthogonal pairs of counter-propagating laser beams of appropriate circular polarizations. However, the space-consuming optical systems for delivering the laser beams and the required polarization optics to produce correct circular polarization states are a significant obstacle for realizing compact and robust system. Thus, some new variations of the MOT such as pyramid MOT (*22*), grating MOT (*23*) and prism MOT (*24*) have been the most popular choices for fully integrated cold atom quantum experiments and devices. The single-beam geometry of these MOTs greatly strengthens the robustness and stability of the devices by simplifying optical delivery of the laser beams and eliminating relative fluctuations in laser power and polarization between different laser beams, which is a crucial task in a conventional MOT configuration. Despite many advances, these variations compromise between the robustness and the performance as they typically create less symmetric illumination of the capture region for the quantum ensemble leading to deformations in the shape of the cloud (*22, 25, 26*) and compromise on the efficiency of delivery of the optical power which is a serious limitation in realizing quantum sensors that meet the demanding size, power and cost constraints of commercial or space applications.

In this paper, we propose a new metasurface approach to address the above issues. The metasurface is designed to diffract a single incident laser beam into five beams with predefined directions and circular polarization states (Fig. 1). With the assistance of five mirrors, all beams intersect at the center of a quadrupole magnetic field, where the wave vectors of laser beams sum to zero to satisfy the three-dimensional cooling and trapping condition. By utilizing the metasurface optical chip in the MOT system, one can replace the conventional optical systems required for manipulating the laser beam, which are typically composed of large and complex arrays of optical elements such as lenses, prisms, polarization converters, etc.

## Results
**Design, Fabrication and Characterization of the Metasurface Optical Chip**
Here the dielectric metasurface optical chip is designed based on the concept of the Pancharatnam-Berry (P-B) phase (*27–29*), also named geometric phase, which only depends on the orientation angle of the anisotropic meta-atoms that act as local half-wave plates. Therefore, the circularly polarized incident light passing through the metasurface is converted into the opposite circular polarization state and acquires an spatially variant phase profile, $\varphi(x, y) = 2\sigma\theta(x, y)$, where $\theta(x, y)$ is the spatial orientation of the meta-atom at position $(x, y)$ and represents the left- and right-circular polarizations (LCP and RCP) of the incident light. The detailed design method of the metasurface optical chip is discussed in the Materials and Methods. In short, we use amorphous silicon nanofins on a glass substrate as the meta-atoms. The specific geometrical parameters of the meta-atom depicted in Fig. 2A are designed by using numerical simulation software (FDTD solver, Lumerical Inc.) to maximize the cross-polarization (LCP/RCP to RCP/LCP) conversion efficiency to about 91% at the wavelength of 780 nm (Fig. 2B). The calculated phase profile for generating the desired diffracted beams is shown in Fig. 2C. The scanning electron microscopy (SEM) image of the metasurface optical chip consisting of an array of meta-atoms is shown in Fig. 2D. To prove the concept of metasurface based MOT, we designed a metasurface optical chip of 599.4 μm × 599.4 μm in size. In principle, it is possible to make metasurface optical chips with inch sizes depending on the experimental requirements.

A detailed characterization of the optical functionalities of the metasurface optical chip is performed using a laser with LCP polarization normally incident on the device. Four beams deflect towards the ±$x$-axis and the ±$y$-axis with angles of 22.5° as shown in Fig. 3A. Here the polarization purity of each first-order diffracted laser beam, characterized by the RCP percentage, reaches around 99%, as shown in Fig. 3B. For the central beam, its power occupies 83% of the total laser power after transmission, the RCP percentage is around 6.7%, which means the central beam has elliptical polarization. In the following MOT experiment, a circular polarizer will be used to filter out the RCP components. The purity of RCP components of the four laser beams at east (E), west (W), central (C), north (N) and south (S) directions is around 99%, while the power differences between the four beams are within 5% (See the Supplementary Materials). The measured total diffraction efficiency, i. e. the proportion of converting the incident LCP state to the RCP state after the metasurface optical chip, is about 22%, which is far below the theoretically predicted value but can be further improved in future by optimizing the nanofabrication processes. The measurements show that all the laser beams have a Gaussian profile, which is important to the performance of MOT for generating a symmetrical radiation pressure and cloud distribution (Fig. 3B). It should be noted that the handedness of four diffracted laser beams is opposite to the one in the center because of field directions of magnetic fields are opposite to each other on the radial and the axial directions (Fig. S4D).

**Performance of the Metasurface based MOT**
The performance of the metasurface based MOT is characterized by the attainable number and the temperature of trapped atoms. The atom number trapped in the MOT is measured at different laser detuning frequencies and coil currents, as shown in Fig. 4A. Approximately $10^7$ atoms are captured by using the optimized parameters, i.e. a detuning -10 MHz and a coil current 4.4 A, which produces a magnetic gradient 16 G/cm. The measured atom number approaches the theoretical limit of ~ $10^7$, corresponding to the laser diameter of ~ 5 mm and the detuning about $2\gamma$ where $\gamma$ ~ 6MHz for rubidium (*30*). A higher atom number can be achieved by enlarging the laser beam size, and is here limited by the metasurface chip size.

The temperature of atom cloud can be deduced from the sequential images of free expansion of the cloud after being released from the metasurface MOT when the quadrupole magnetic field is off. The detail of the procedures is described in the Supplementary Materials. Fig. 4B shows the sequential images of atomic cloud released from the MOT. Fig. 4C shows the evolution of the radii of the cloud in the time-of-flight measurement, in which the temporal variation of the cloud size is plotted as the blue trace. Using a Gaussian fitting model, the temperature found from the expansion are determined as 764.1 μK and 2300.0 μK in the axial and radial direction, respectively. Subsequently, three pairs of Helmholtz coils are used to compensate the ambient magnetic field at the center of the quadrupole magnetic field in the experiment. Then, the magnetic field of the MOT is turned off and a polarization gradient cooling is applied for 10 milli-seconds to further cool the atoms to the sub-Doppler temperature. Fig. 4D shows that the expansion of the atom cloud after this further cooling step is much slower, corresponding to a temperature of 35.2 µK and 36.9 µK in the axial and radial direction, respectively. This significant reduction of the temperature can be attributed to the symmetrical and well-balanced radiation pressures generated by the metasurface, which provides a promising candidate for compact cold atom source.

**Discussion**

In summary, we have provided the first demonstration of the use of metasurface optical chips as a new approach for the generation of cold atoms and assessed their initial performance through realizing initial atom numbers and temperatures commensurate with quantum sensing. We achieve temperatures comparable to what one would get with similar setups and boundary conditions in standard MOT systems. Our experiment does not show any limitations of the temperature imposed by the metasurface technology. Without further optimization, the temperature achieved with the metasurface optical chip is already sufficient for operating compact high-bandwidth atom interferometers and similar to the temperatures used when loading magnetic traps for further evaporative cooling to BEC. A pathway to further reductions of temperature e. g. for applications in atomic fountains would be by applying similar optimization techniques as used in standard MOTs for such applications, i. e. optimizing polarization balance, magnetic field compensation and using larger beam diameters. This work provides a new and highly promising approach for the realization of compact and future commercial quantum sensors, in particular through enabling more compact and lower power systems. Furthermore, it opens a new direction in the application of metasurface for drastically improving the delivery of atom optics or optical lattices changing the capabilities of a wide range of quantum sensing modalities, and having broad application in quantum metrology, quantum information processing and atomic physics.

**Materials and Methods**

**Design of Metasurface Optical Chip**

In principle, the circularly polarized laser normally incident on the metasurface optical chip can be divided into five beams with the same intensity and circular polarizations. One of them propagates along the incident direction, while the other four beams are deflected towards ± $x$-axis and ± $y$-axis with an angle of 22.5° with respect to the incident beam, respectively. To design the metasurface optical chip, we use a reciprocal process and superpose the five output beams together at the metasurface plane ($z$ = 0 plane). In this way, the required phase profile of the Pancharatnam-Berry phase type metasurface is given by:

$$\Phi(x,y) = \arg\left(\sum_{n=0}^{4} A_n \cdot e^{-i\cdot \mathbf{k}_n \cdot \mathbf{r}_n}\right).$$

Where the arg function returns the argument of the complex amplitude, $n$ = 0, 1, 2, 3, 4 represent the five output beams, $A_n$ is the amplitude of the electric field of light which equal to 1 for all $n$. $\mathbf{k}_n$ and $\mathbf{r}_n$ are the wavevector and position vector of each beam. $|\mathbf{k}_n| = 2\pi/\lambda_0$, where $\lambda_0$ = 780 nm is the wavelength of light in free space. According to the geometrical parameters of the amorphous silicon meta-atom, a metasurface optical chip with size of 599.4 µm × 599.4 µm is designed.

**Polarization Measurement of the Laser Beams**

The setup to measure the polarization of the laser beams after the metasurface optical chip is shown in Fig. S3. The linear polarization of the incident laser is improved further by a Glan-Thomson polarizer (extinction ratio = 100000:1). The laser can be configured in the LCP or RCP by changing the angles of the optical axis between the Glan-Thomson linear polarizer (LP1) and the quarter-wave plate (QWP1). Then the laser is focused on the metasurface optical chip after passing a lens (f = 200 mm). The output after the chip is transmitted through the QWP2 and LP2. By changing the alignment of the optical axis of the QWP2, we can choose the LCP or RCP constituents in the laser passing through the LP2. The results of the measurement are shown in Table S1.

**MOT Apparatus and Procedures**
The MOT apparatus (Fig. S5) comprises a light distribution frame combining with anti-Helmholtz coils (AHC) producing the quadrupole magnetic field and a vacuum chamber containing the rubidium atoms. There are extra three pairs of Helmholtz coils installed around the system to compensate for the ambient magnetic field. The distributing frame is made by Plexiglas and four mirrors are attached on angled mounts to direct the beams. The AHC are mounted on the rails to align the center of the quadrupole field to the center of the intersection region of optical beams. The coil current of 4.4 A produces a 16 G/cm magnetic field gradient which is desirable for the operation of the Rubidium (Rb) MOT. The main vacuum chamber is an AR-coated glass cell whose dimension is 35 mm × 24 mm × 60 mm and is pumped by a 2 L/s ion pump. The rubidium atoms are produced by heating a rubidium dispenser with an electric current flowing through. The background pressure is about $2 \times 10^{-9}$ mbar as the dispenser is cool and raised to $2 \times 10^{-8}$ mbar when the experiment is running.

**Acknowledgments:** The authors thank the fruitful discussions with X. Li, L. Huang and J. Rho at the early stage of this project. L. Z. would like to thank H. Zhang and X. Zhang for preparing the Figures.

**Funding**: G. L. is financially supported by National Natural Science Foundation of China (91950114 and 11774145), Guangdong Provincial Innovation and Entrepreneurship Project (2017ZT07C071) and Shenzhen DRC project [2018]1433. M. W., Y.-H. L., M. H. and K. B. are supported by Engineering and Physical Sciences Research Council (EPSRC, EP/M013294/1 and EP/T001046/1). T. Z. and S.Z. acknowledge the funding from the European Union's Horizon 2020 research and innovation program (724306 and 648783). L. Z. is supported by Young Elite Scientists Sponsorship Program by CAST (2018QNRC001).

**Author contributions**: G. L., Y.-H. L. and L. Z. conceived the idea. X. L. and G. L. carried out the simulations. B. S., C. S. and T. Z. fabricated the metasurface. L. Z., K. L., Y. T. and J. D. measured the polarization states of the metasurface optical chip. Y.-H. L. and M. W. conducted the measurement of performance of the cold atom device. M. H. and K. B. contributed to the scientific discussion, the implementation of the cold atom device. Y.-H. L. and G. L. supervised the project. All authors contributed to the data analysis and manuscript writing. L. Z., X. L., B. S. and M. W. contribute equally.

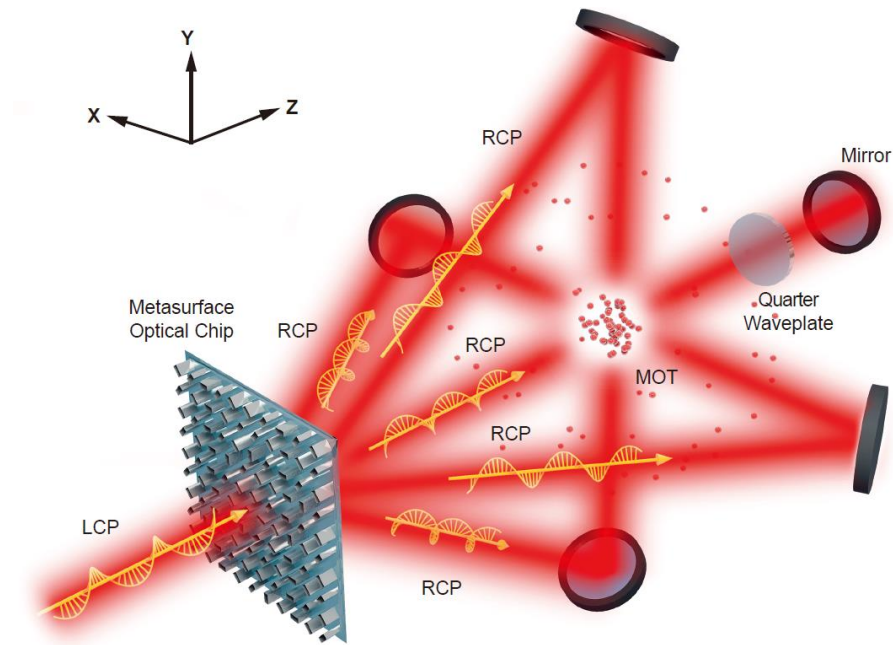

**Fig. 1. Schematic of the cold atom device with the dielectric metasurface optical chip.** A Left-circularly polarized (LCP) light passing through the medium is diffracted into five right-circularly polarized (RCP) beams, respectively. After the reflection on the mirrors, all the beams are overlapped with the appropriate polarization of the MOT. Atoms are trapped near the zero of a quadrupole magnetic field located within the beam overlapping region. The trapping volume in the metasurface based MOT has the same shape as that in the six-beam MOT.

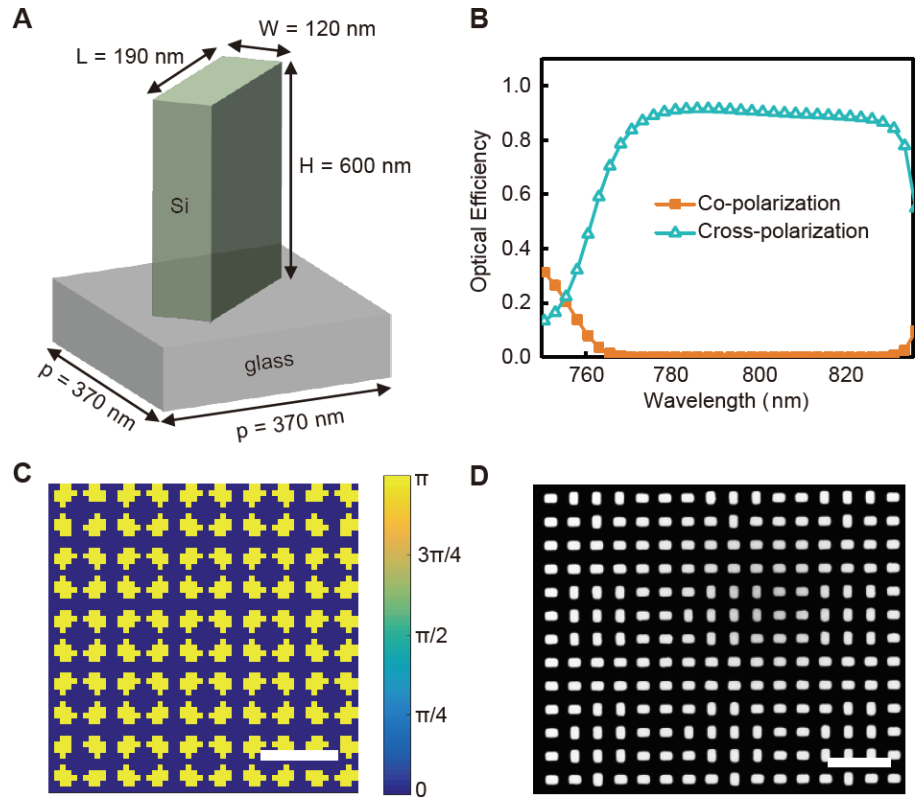

**Fig. 2. Design and fabrication of the metasurface optical chip.** (**A**) The geometric configuration of the meta-atom with period p = 370 nm, meta-atom length L = 190 nm, width W =120 nm and height H = 600 nm. (**B**) Simulated cross-polarization and co-polarization conversion efficiency for different wavelengths of the incident light. (**C**) The calculated phase profile of the metasurface optical chip. Scale bar: 5 μm. (**D**) SEM image of the fabricated metasurface optical chip (partial view), scale bar: 1μm.

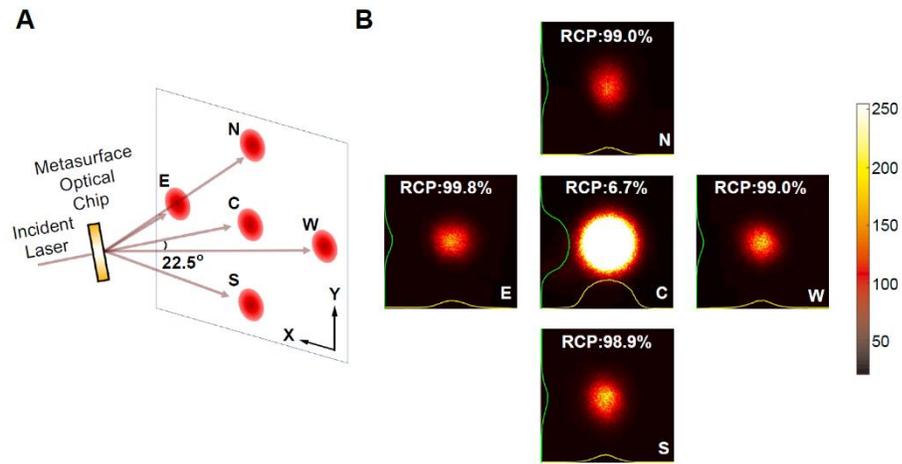

**Fig. 3. Measurement of the optical performance of the metasurface optical chip.** (**A**) Schematic of the spatial distribution of the transimitted laser beams after the metasurface optical chip. The laser beams are labelled depending on the location on the projection screen by east (E), west (W), central (C), north (N) and south (S), respectively. (**B**) Intensity profile of the laser beams captured by a CCD camera after the metasurface optical chip. The incident laser beams are right circularly polarized. All the laser intensity profiles are plotted with the same scale of the colorbar. The laser intensity integral along the *x*- and *y*- axis are plot as the green and yellow curves in each image. The corresponding RCP percentage of each beam is shown on the top of each image.

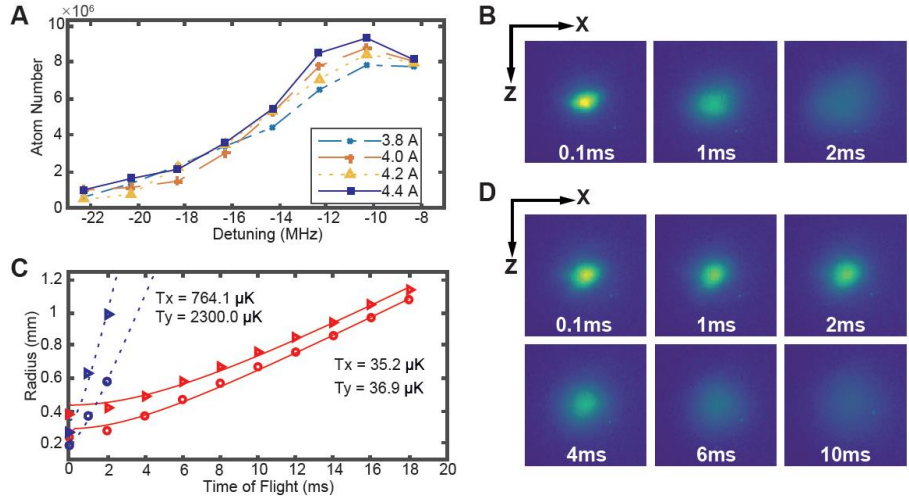

**Fig. 4. Performance of the metasurface based MOT.** (**A**) Variation of the number of atoms trapped in the metasurface based MOT with the laser detuning. For each curve, a fixed intensity was used. Data points are the average of five runs, resulting in statistical uncertainties much smaller than the plot markers. The atom number peaks at a detuning around 10 MHz below resonance, nearly two linewidths away of the $^{87}$Rb transition. The atom number reaches a value close to $10^7$ at a coil current of 4.4 A. (**B**) Absorption images of the $^{87}$Rb cloud at expansion times after MOT. (**C**) The fitted $1/\sqrt{e}$ (68% probability) radii of the $^{87}$Rb cloud in the axial (circles) and radial (triangles) directions versus the expansion time $t$. The expansion of the cloud after the MOT (blue) and optical molasses (red) is fitted in dotted and solid curves, respectively. (**D**) Absorption images of the $^{87}$Rb cloud at expansion times after optical molasses.